\documentclass[12pt]{article}

\usepackage{graphicx}
\usepackage{algorithm,algorithmicx,algpseudocode}
\usepackage{enumerate}
\usepackage{amsfonts}
\usepackage{latexsym,amsmath}

\usepackage{slashbox}
\usepackage{amssymb,array}
\begin{document}
\title{\large \bf Cryptanalysis of Wu and Xu's authentication scheme  for Telecare Medicine Information Systems}

\author{\small Dheerendra Mishra\thanks{Corresponding author. E-mail:~{dheerendra@maths.iitkgp.ernet.in} }, and Sourav Mukhopadhyay\\
\small Department of Mathematics,\\
\small Indian Institute of Technology Kharagpur,\\
\small  Kharagpur 721302, India\\}

\date{}
 \maketitle

\begin{abstract}
Remote user authentication is desirable for a Telecare medicine information system (TMIS) to verify the correctness of remote users. In 2013, Jiang et al. proposed privacy preserving authentication scheme for TMIS. Recently, Wu and Xu analyzed Jiang's scheme and identify serious security flaws in their scheme, namely, user impersonation attack, DoS attack and off-line password guessing attack. In this article, we analyze Wu and Xu's scheme and show that their scheme is also vulnerable to off-line password guessing attack and does not protect user anonymity. Moreover, we identify the inefficiency of incorrect input detection of the login phase in Wu and Xu's scheme, where the smart card executes the login session in-spite of wrong input.

\end{abstract}
\textbf{keywords:} {Telecare medicine information system; Smart card; Remote user authentication; Biometric; Password.}

\section{Introduction}{\label{intro}}
Telecare medicine information system (TMIS) provides efficient medical services for remote user. In these online services user and medical server communicate via public channel, which increases security and privacy threat. Although, password based authentication schemes provides efficient remote user authentication mechanism for TMIS, where user and server can mutually authenticate each other and establish a secure session by drawing unique session key for each session. In recent years many password based authentication schemes have been proposed for TMIS~\cite{cao2013improved,chen2012efficient,debiao2012more,lin2013security,wei2012improved,wu2012secure,xie2013robust,zhu2012efficient}, which tries to provide two factor authentication.

In 2010, Wu et al.~\cite{wu2012secure} presented an efficient authentication scheme for TMIS. Their scheme is better than the before proposed schemes for low computing devices, which added pre-computing phase.  In this phase, user performs exponential operation and then stores the calculated values into the storage device, where user can extract them from the device when required. However, Debiao et al.~\cite{debiao2012more} demonstrated that Wu et al.'s scheme does not resist the impersonation attack to the insider's attack and introduced an enhanced scheme that eliminates the drawbacks of Wu et al.'s scheme. Additionally, it performs better than Wu et al.'s scheme.  They also claimed that their scheme is more appropriate to low power mobile devices for TMIS. In 2012, Wei et al.~\cite{wei2012improved} identified that both Wu et al.'s and Debiao et al.'s schemes are failed to meet two-factor authentication, although an efficient smart card based password authentication schemes should support two-factor authentication. They also presented an improved authentication scheme for TMIS and claimed that the improved scheme is efficient and achieved two-factor authentication. However,  Zhu~\cite{zhu2012efficient} demonstrated that Wei et al.'s scheme is vulnerable to an off-line password guessing attack using stolen smart card.  He also presented an improved authentication scheme for TMIS and claimed that his scheme could overcome the weaknesses of Wei et al.'s scheme. In 2012, Chen et al.~\cite{chen2012efficient} also presented an efficient and secure dynamic ID-based authentication scheme for TMIS. In 2013, Jiang et al.~\cite{jiang2013privacy} showed that does not achieve user anonymity and presented an improved scheme. Recently, Wu and Xu demonstrated that Jiang et al.'s scheme is vulnerable to vulnerable to user impersonation attack, off-line password guessing attack and DoS attack. They also presented an improved scheme for TMIS and claimed that their scheme overcomes the disadvantages of Jiang et al.'s scheme and is suitable for TMIS.

In this article, we analyze the Wu and Xu's password based remote user authentication scheme for TMIS.  We noted that Wu and Xu's scheme does not present efficient login and user-friendly password change phase. Inefficiency of login phase in incorrect input detection causes extra communication and computation overhead, as smart card executes the login session in-spite of wrong identity and password input. Moreover, Wu and Xu's scheme is vulnerable to off-line password guessing attacks and does not protect user anonymity.

The rest of the paper is organized as follows: the brief review of Wu and Xu's scheme is presented in Section \ref{review}. In Section \ref{crypt}, we demonstrate the weaknesses of Wu and Xu's scheme. The conclusion is drawn  in Section \ref{conclusion}.

\renewcommand{\labelitemi}{$ $}
\section{Review of Wu and Xu's Scheme}\label{review}

We will define all the notation

\begin{table}[ht]
  \caption{Meaning of symbols used throughout the paper}\label{t1}
\begin{tabular}{l|l}
\hline
Notation & Descryption\\
\hline
$U_i$ & User $i$\\
$S$ & A trustworthy medical server\\
$E$ & Attacker/adversory\\

$ID_i$ & Unique identity of user $i$\\
$PW_i$  & Unique password of user $i$\\
$N$     & Registration time of user\\
$T_i$ & Timestamp generated by $U_i$\\
$T_s$ & Timestamp generated by $S$\\
$sk$ & Session key\\
$x$  & Master key of $S$\\
$E_{(x)}(M)$ & Symmetric key encryption of message $M$ using key $ x$\\

$D_{(x)}(M)$ & Symmetric key decryption of message $M$ using key $x$\\

$h(\cdot)$ &  A collision free one-way hash function\\

$\oplus$ & XOR\\
$||$ & String concatenation operation\\

 \hline
\end{tabular}
\end{table}

In 2013, Wu and Xu~\cite{wu2013security}  proposed an improvement of Jiang et al.'s~\cite{jiang2013privacy} privacy enhanced password-based authentication scheme for TMIS. This comprises following five phases:
\begin{itemize}
  \item Registration
  \item Login
  \item Authentication
  \item Password change
  \item Lost smart card revocation
\end{itemize}

In this section, we will briefly discuss registration, Login and authentication phases of Wu and Xu's scheme, as we use only these phases in our discussion. For more detailed study of Wu and Xu's scheme, one can refer~\cite{wu2013security}.

\subsection{Registration Phase}
In registration phase, a new user registers to the server and achieves the smart card as follows:

\begin{description}

\item [\bf Step 1.] $U_i$ chooses his identity $ID_i$ and password $PW_i$, and generates a random number $r_i$. He computes $RPW_i = h(r_i||PW_i)$, then sends the message $<ID_i, RPW_i>$ to $S$ via secure channel.

\item[\bf Step 2.] Upon receiving the registration request, $S$ verifies the validity of $ID_i$. If verification does not hold, it terminates the session. Otherwise, if $U_i$ is a new user, $S$ computes $J_i = h(x||ID_i||N)$, $L_i = J_i\oplus RPW_i$ and $e_i = h(x)\oplus h(RPW_i||ID_i)$, where $N= 0$ for new user, otherwise $N = N+1$. Then, $S$ embeds $\{L_i, e_i, h(\cdot), E_{key}(\cdot), D_{key}(\cdot)\}$ into smart card and issues the smart card to $S$. Moreover, $S$ also includes ($ID_i, N)$ into its database.

\item [\bf Step 3.] Upon receiving the smart card, $U_i$ stores $r_i$ into the smart card. 

\end{description}
%



\subsection{Login Phase}
When a user wishes to login to the server, he first inserts his smart card into the card reader and inputs the identity $ID_i$ and password $PW_i$, then login process executes as follows:

\begin{description}

\item[\bf Step 1.] Smart card computes $RPW_i = h(r_i||PW_i)$, $J_i = L_i\oplus RPW_i$ and $AID = e_i\oplus h(RPW_i||ID_i)\oplus h(T_i)\oplus ID_i$, $B_1 = e_i\oplus h(RPW_i||ID_i)\oplus T_i$, $V_i = h(T_i||J_i)$ and $C_1 =  E_{h(T_i)}(AID_i||T_i||V_i)$.

\item[\bf Step 2.] Smart card sends the login message $<B_1, C_1>$ to $S$.

\end{description}


\subsection{Authentication Phase}
User and server performs the following steps to mutually authenticate each other:

\begin{description}

\item [\bf Step 1.] Upon receiving the message $<B_1, C_1>$, $S$ achieves $T_i' = B_1\oplus h(x)$ and verifies the validity of timestamp. If timestamp is invalid, it terminates the session. Otherwise, $S$ computes $h(T_i')$ and decrypts $C_1$ with $h(T_i')$ and achieves $AID_i', T_i''$ and $V_i'$. Then $S$ verifies the condition $T_i' =? ~ T_i''$. If the condition does not hold, it terminates the session. Otherwise, $S$ computes $ID_i' = AID'\oplus h(x)\oplus h(T_i')$, then searches $ID_i'$ in its database. If search fails, it terminates the session. Otherwise, $S$ achieves $N$ from its database and computes $J_i' = h(x||ID_i'||N)$, then verifies $V_i' = h(T_i'||J_i')$. If verification does not hold, it terminates the session. Otherwise, it runs {\em Step 2}.

\item[\bf Step 2.] $S$ computes $B_2 = h(x)\oplus T_s$, $C_2 = E_{h(T_s)}(V_i'||T_s)$ and $sk = h(J_i'||T_i'||T_s||$ $ID_i')$. Then, $S$ sends the message $<B_2, C_2>$ to $U$.

\item[\bf Step 3.] Upon receiving the message  $<B_2, C_2>$, smart card computes $T_s' = B_2\oplus e_i\oplus h(RPW_i||ID_i)$. Then, it verifies the validity of $T_s'$. If verification holds, it decrypts $C_2$ using $h(T_s')$, {\em i.e.}, $D_{h(T_s')}(C_2) = V_i''||T_s''$. Finally, it verifies $V_i'' =? ~V_i$ and $T_i'' =? ~T_i'$. If verification does not succeed, it ends the session. Otherwise, it computes the session key $sk = h(J_i||T_i||T_s'||ID_i)$.

\end{description}





\renewcommand{\labelitemi}{$-$}
\section{Cryptanalysis of Wu and XU's Scheme}\label{crypt}

In this section, we show that Wu and XU's Scheme~\cite{yan2013secure} does not efficient login and user-friendly password change phase. Moreover, their scheme is vulnerable to off-line password guessing attack and does not protect user's anonymity. This analysis is based on the on the following assumptions, which are stated in the literature~~\cite{brier2004correlation,eisenbarth2008power,kocher1999differential,messerges2002examining,rankl2010smart,boyd2003protocols,xu2009improved}.

\begin{itemize}
  \item An adversary is able to extract the information from the smart card or mobile device.
  \item An adversary is able to  eavesdrop all the messages between user and server, which transmits via public channel. Moreover, is able to modify, delete and resend all the messages,  and can also reroute any message to any other principal.
  \item An adversary may be a legitimate user or an outsider and can register as a legitimate user and achieve the smart card.
\end{itemize}

Due to above mentioned assumptions, an adversary can achieve the parameters $\{L_i = J_i\oplus RPW_i, e_i = h(x)\oplus h(RPW_i||ID_i), r_i, h(\cdot), E_{key}(\cdot), D_{key}(\cdot)\}$ from the smart card   and can intercept and record the messages $<B_1, C_1>$ and $<B_2, C_2>$, which transmits via public channel. 

\renewcommand{\labelitemi}{$\bullet$}
\subsection{Off-line password guessing attack}
The password guessing attack is the most common attack on password based authentication protocols. Although an efficient authentication scheme should resist password guessing attack. When we investigated Wu and Xu's scheme, we identify that their scheme does not withstand Off-line password guessing attack. This is clear from the following facts:

 \begin{description}
 \item[\bf Step p1.] An attacker achieve the stored parameters $\{L_i, e_i, r_i, h(\cdot), E_{key}(\cdot),$ $ D_{key}(\cdot)\}$ from the smart card, where $L_i = J_i\oplus RPW_i$ and  $e_i = h(x)\oplus h(RPW_i||ID_i)$.

  \item[\bf Step p2.] An attacker can intercept user's message $<B_1, C_1>$  and record it, where $B_1 = e_i\oplus h(RPW_i||ID_i)\oplus T_i = h(x)\oplus T_i$, $V_i = h(T_i||J_i)$ and $C_1 =  E_{h(T_i)}(AID_i||T_i||V_i)$.

 \item [\bf Step p3.] An attacker $E$ may be a legitimate user, then by using his identity $ID_E$ and password $PW_E$, he can achieve the smart card with the personalized parameters $\{L_E, e_E, r_E, h(\cdot), E_{key}(\cdot), D_{key}(\cdot)\}$, where $RPW_E = h(r_E||PW_E)$, $L_E = J_E\oplus RPW_E$ and  $e_E = h(x)\oplus h(RPW_E||$ $ID_E)$ by registering to $S$. With the help of achieved parameters, attacker can achieve $h(x)$ as follows:
     \begin{itemize}
       \item  Compute $RPW_E = h(r_E||PW_E)$.
       \item   Compute $h(x) = e_E \oplus h(RPW_E||ID_E)$.

     \end{itemize}
\item[\bf Step p4.] Attacker guesses user $U_i$ password $PW^*$ and compute $RPW_i^* = h(r_i||PW_i^*)$ and $J_i^* = L_i\oplus RPW_i^*$, then achieves $T_i$ from $B_i$ as $T_i = B_i\oplus h(x)$. Then, attacker decrypts $C_i$ using $h(T_i)$ and achieves $(AID_i||T_i||V_i)$, as $D_{h(T_i)}(C_1) =  D_{h(T_i)}(E_{h(T_i)}(AID_i||T_i||V_i)) = (AID_i||T_i||V_i)$. Finally, an attacker verifies the guessed password using the condition $V_i =?~ h(T_i||J_i^*)$.

\item[\bf Step p5.] If the verification succeeds, then the password guessing succeeds. Otherwise, the attacker repeats {\em Step p4}.

  \end{description}

\subsection{Flow in privacy protection}
Wu and Xu's claimed that their scheme protect user anonymity. However, we identify that an attacker can easily track the user in Wu and Xu's scheme by performing the following steps:

\begin{itemize}
  \item  An attacker can intercept the user $U_i$'s login message and achieve $<B_1, C_1>$.
  \item An attacker can achieve $h(x)$, as discussed in {\em Step p3}. An attacker can also achieve $T_i$ from $B_1$ as $B_1 = h(x)\oplus T_i$.
  \item  With the help of achieved $T_i$, an attacker can decrypt $C_i$, {\em i.e.},  $D_{h(T_i)}(C_1) =  D_{h(T_i)}(E_{h(T_i)}(AID_i||T_i||V_i)) = (AID_i||T_i||V_i)$, where $AID = e_i\oplus h(RPW_i||ID_i)\oplus h(T_i)\oplus ID_i$.
  \item An attacker can achieve user identity $ID_i$ using $AID_i$, $h(x)$ and $h(T_i)$ as follows:

  \begin{eqnarray*}
    AID_i\oplus h(x)\oplus h(T_i) &=& e_i\oplus h(RPW_i||ID_i)\oplus h(T_i)\oplus ID_i \oplus h(x)\oplus h(T_i) \\
     &=& h(x)\oplus h(RPW_i||ID_i)\oplus h(RPW_i||ID_i)\oplus ID_i \oplus h(x)\\
     &=& ID_i
  \end{eqnarray*}
\end{itemize}

The above discussion shows that, an adversary can achieve user identity from the intercepted message. This shows that Wu and Xu's scheme is failed to prove its claim of user anonymity.

\subsection{Flaws in login phase}
In Wu and Xu's scheme, smart card does not verify the correctness of input, {\em i.e.}, correctness of identity and password in login phase. However, a user may enter wrong password or identity due to mistake. If a user enters wrong input, then the following cases are possible:

\noindent\textbf{Case 1:} If a user inputs wrong password $PW_i^*$ due to mistake. The smart card still executes the login phase as follows:

\begin{itemize}

\item Without verifying the correctness of input, the smart card computes $RPW_i^* = h(r_i||PW_i^*) \neq h(r_i||PW_i) = RPW_i$ as $PW_i\neq PW_i^*$ and $J_i^* = L_i\oplus RPW_i^*$.

\item Then, the smart card computes  $AID_i^*$ and $B_i^*$ as follows:

 \begin{eqnarray*}
      AID_i^* &=& e_i\oplus h(RPW_i^*||ID_i)\oplus h(T_i)\oplus ID_i\\
              &=& h(x)\oplus h(RPW_i||ID_i)\oplus h(RPW_i^*||ID_i)\oplus h(T_i)\oplus ID_i \\
       &\neq& h(x) \oplus h(T_i)\oplus ID_i, ~\texttt{ as} ~ h(RPW_i||ID_i)\neq h(RPW_i^*||ID_i)\\
     B_1^*  &=& e_i\oplus h(RPW_i^*||ID_i)\oplus T_i\\
       &=&  h(x)\oplus h(RPW_i||ID_i)\oplus h(RPW_i^*||ID_i)\oplus T_i\\
       &\neq& h(x)\oplus T_i  ~\texttt{ as} ~ h(RPW_i^*||ID_i)\neq h(RPW_i||ID_i)
    \end{eqnarray*}

\item   Smart card also computes
     $V_i^* = h(T_i||J_i^*)$ and $C_1^* =  E_{h(T_i)}(AID_i^*||T_i||V_i)$.
     Finally, it sends the login message $<B_1^*, C_1^*>$ to $S$.

\item Upon receiving the message $<B_1^*, C_1^*>$, $S$ computes $T_i^* = B_1^*\oplus h(x) \neq T_i$, as $B_1^*\neq h(x)\oplus T_i$. The verification of timestamp does not holds. Moreover, when $S$ try to decrypt $C_i^*$, the procedure fails, as $S$ could not achieve correct $T_i$.

\item It is clear that in case of wrong password input, the verification does not hold at server end. Thus, $S$ terminates the session.

\end{itemize}

\noindent\textbf{Case 2:} If a user inputs the incorrect identity $ID_i^*$, the smart card does not verify the correctness of identity and executes the session as follows:

\begin{itemize}

\item Without verifying the correctness of input, the smart card computes $RPW_i = h(r_i||PW_i)$, $J_i = L_i\oplus RPW_i$. Then, smart card computes the following values:

    \begin{eqnarray*}
      AID_i^* &=& e_i\oplus h(RPW_i||ID_i^*)\oplus h(T_i)\oplus ID_i^*\\
               &=& h(x)\oplus h(RPW_i||ID_i)\oplus h(RPW_i||ID_i^*)\oplus h(T_i)\oplus ID_i^* \\
       &\neq& h(x) \oplus h(T_i)\oplus ID_i, ~\texttt{ as} ~  ID_i\neq ID_i^*\\
     B_1^*  &=& e_i\oplus h(RPW_i||ID_i^*)\oplus T_i\\
       &=&  h(x)\oplus h(RPW_i||ID_i)\oplus h(RPW_i||ID_i^*)\oplus T_i\\
       &\neq& h(x)\oplus T_i  ~\texttt{ as} ~ h(RPW_i||ID_i)\neq h(RPW_i||ID_i^*)
    \end{eqnarray*}

 \item  The smart card also computes
     $V_i = h(T_i||J_i)$ and $C_1^* =  E_{h(T_i)}(AID_i^*||T_i||V_i)$.
     Finally, it sends the login message $<B_1^*, C_1^*>$ to $S$.

\item Upon receiving the message $<B_1^*, C_1^*>$, $S$ computes $T_i^* = B_1^*\oplus h(x) \neq T_i$, as $B_1^*\neq h(x)\oplus T_i$. The verification of timestamp does not holds. Moreover, when $S$ tries to decrypt $C_i^*$, the procedure fails, as $S$ could not achieve correct $T_i$.

\item It is clear that in case if user inputs wrong identity, the verification does not hold at server end. Thus, $S$ terminates the session.

\end{itemize}

It is clear from the above discussion that in case of wrong input smart card execute the session. Moreover, due to inefficiency of incorrect input detection, sever may flooded with fraud inputs. This will decrease the efficiency of the system.

\subsection{Unfriendly password changes phase}
In the proposed scheme, the user has to establish an authorized session with the server, {\em i.e.}, the user can not change his password independently. A user has to depend on server to change his password. However, password change phase should be user-friendly where a user can change his password without server assistance.

\section{Conclusion}\label{conclusion}
The password based authentication schemes provide efficient and scalable solutions for remote user authentication. Recently, Jiang et al. proposed a password based remote user authentication scheme for the TMIS. In more recent work of Wu and Xu identified the vulnerability of Jiang et al.'s scheme to resist denial-of-service attack, off-line password guessing attack and user impersonation attack. Further, they proposed an improved scheme to overcome the disadvantage  of Jiang et al.'s scheme. In this paper, we analyzed Wu and Xu's scheme and identified that their scheme is vulnerable to off-line password guessing attack and fails to present efficient login phase phase, where smart card can identify the incorrect input. Moreover, we showed that Wu and Xu's scheme does not protect user anonymity.

%

\end{document}